\documentclass[ preprintnumbers, amsmath, amssymb, twocolumn,showpacs]{revtex4}

\usepackage{graphicx}
\usepackage{dcolumn}
\usepackage{bm}
\usepackage{subfigure}

\begin{document}

\preprint{...}

\title{Odd-even effect of melting finite polymer film on square lattice}

\author{Tieyan Si}
\affiliation{Academy of Fundamental and Interdisciplinary Sciences,
Harbin Institute of Technology, Harbin, 150080, China}

\date{\today}

\begin{abstract}

Two dimensional film system bears many exotic thermodynamics
behaviors. We proposed a mathematical physics model to explore how
the melting temperature of a two dimensional mathematical dimer film
depends on the odd-eveness of the finite width of dimer film. A weak
external bond between dimers is introduced into the classical dimer
model in this dimer film. We derived a general equation of melting
temperature and applied it for computing the melting temperature of
a dimer film covering a finite square lattice. The melting
temperature is proportional to the external bonding energy that we
assume it binds neighboring dimers together and proportional to the
inverse of entropy per site. Further more, it shows fusing two small
rectangular dimer film with odd number of length into one big
rectangular film gains more entropy than fusing two small rectangles
with even number of length into the same big rectangle. Fusing two
small toruses with even number of length into one big torus reduces
entropy. Fusing two small toruses with odd number of length
increases the entropy. Thus two dimer films with even number of
length repel each other, two dimer films with odd length attract
each other. The odd-even effect is also reflected on the correlation
function of two topologically distinguishable loops in a torus
surface. The entropy of finite system dominates odd-even effect.
This model has straightforward extension to longer polymers and
three dimensional systems.

\end{abstract}


\maketitle


\section{Introduction}

The melting of two dimensional solid film, such as polymer film,
liquid-crystal film, vortex film, was theoretically suggested by
beginning with dissociation of vortex pairs or topological defect
\cite{kosterlitz}. In recent years, a multiple-step melting behavior
of two dimensional liquid-crystal film was experimentally observed
\cite{Chou}. Unlike the three dimensional system, the odd-even
effect in two dimensional system is more subtle and hard to
implement by practical equipment. When chemist study the miscibility
of binary mixtures of several low molar mass nematogens with a main
chain liquid crystal polymer (TPB-x), it was found only when the
length of the flexible alkyl spacers in liquid crystal polymer is
odd, the binary mixtures are miscible. If the length of the flexible
alkyl spacers is even, the mixtures of several low molar mass
nematogens is immiscible \cite{chen}. The odd-even effect is also
illustrated by the melting temperature of n-alkanes $C_{n}H_{2n+2}$
\cite{baeyer}. The melting point of fatty acids with odd number of
chain length are below its neighboring even number of chain length
\cite{baeyer}. When the length of the n-alkanes increases by one
unit from an even number to its next neighboring odd number, the
melting temperature increases by a relatively smaller value compared
with the case that from an odd number to even number
\cite{Boese}\cite{Hagele}. This odd-even effect is diminished when
the chain length becomes longer than 16. The experiments believed it
is the different crystalline structures of the odd and even
oligomers that results in the odd-even effect and anomalous odd-even
effect \cite{chen}\cite{baeyer}\cite{Boese}\cite{Hagele}. A powerful
Monte Carlo computer simulation of n-alkanes reported the odd-even
dependence of the triple point temperature on the length of
n-alkanes molecules \cite{Malanoski} as well as some ordered phases
of a lattice-gas systems of dimers, trimers, and tetramers
\cite{Malanoski2}\cite{Yoon}. There is still no exact numerical
simulation that could matche exactly the experimental data of
melting temperature of three dimensional powder of polymers with
different length.

Inspired by the experimentally observed odd-even phenomena above, we
proposed an exact theoretical model for computing the melting
temperature of polymer film since there is still no such an exact
equations for computing the meting temperature so far. For a
straightforward physical understanding on the odd-even effect of
melting polymer by exactly solvable mathematical physics model, we
study how the melting temperature of a dimer film covering a square
lattice depends on the finite width of the dimer film.

\begin{figure}[htbp]
\centering
\par
\begin{center}
\includegraphics[width=0.32\textwidth]{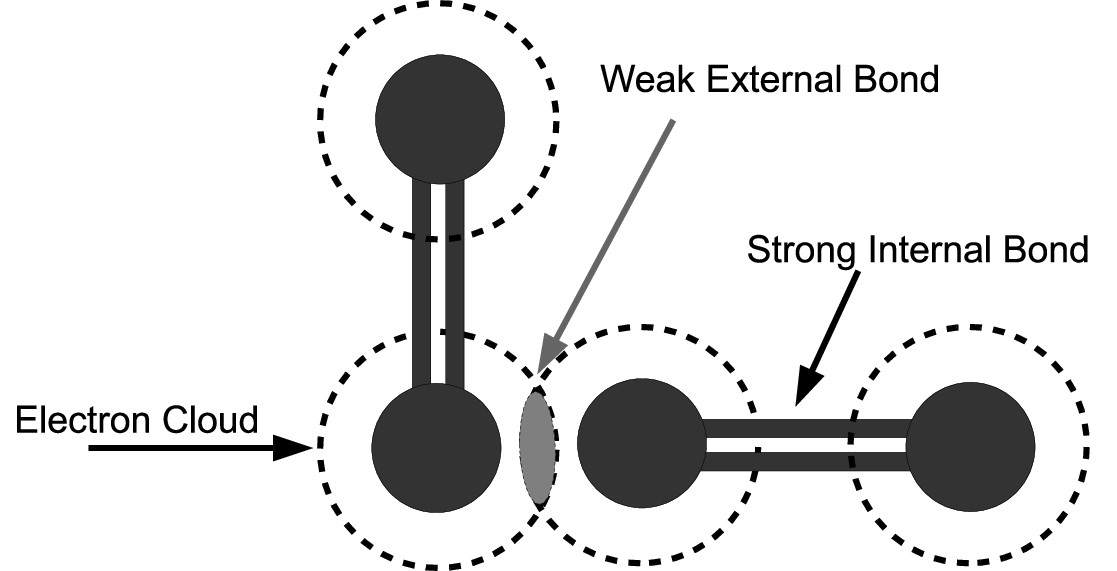}
\end{center}
\caption{\label{bond} We assumed there exists a weak external
bond(the overlapping zone) between dimer molecules due to the
fluctuating overlap of electron cloud. The internal bond is chemical
bond which is a thousand times stronger than the external bond. This
weak external bond could be estimated by the familiar van der Waals
interacton between molecules which decays following $1/r^{6}$. }
\vspace{-0.2cm}
\end{figure}

As there is no report of experimental evidence on the order of phase
transition about melting two dimensional film, we take a naive
assumption that melting two dimensional dimer film is a first order
phase transition. Computing the temperature requires an exact
calculation of internal energy difference and entropy difference.
The internal energy difference is straightforward to calculate. The
entropy difference can be computed by classical dimer model.
Classical dimer model counts how many different ways to cover a
graph by pairing up two neighboring vertex \cite{kenyon}. Kasteleyn
found a method to compute the exact number of dimer covering on
square lattice \cite{Kasteleyn}. The dimer configurations has a
theoretical correspondence with fluctuating string configuration on
two dimensional lattice \cite{Yokoi}. There is no external bonding
between dimers in classical dimer model except that the overlap of
two dimers is forbidden, while here we introduced a weak external
bond that binds neighboring dimers together in this dimer film model
to simulate the melting phase transition(Fig. 1). The odd-even
effect showed by this exact mathematical physics model is different
from that of three dimensional polymer powders. The entropy
variation for fusing two finite films into one or dividing one film
into two also shows odd-even effect. As all know, cell division is a
crucial process for life. While the liquid crystal membrane model
had successfully predicted many properties of the biconcave shape of
red blood cell \cite{ouyang}. The stable existence of the torus of
liquid crystal membrane was first predicted by Helfrich's theory of
lipid blayer membranes\cite{ouyang2} and was confirmed by experiment
later\cite{rudolph}. The second law of thermal dynamics holds both
for liquid crystal film and polymer film. Thus we focus on the
fundamental physical principal of film fusion or division without a
specific molecules. The entropy principal for the odd-even effect of
finite film division is also a fundamental principal for
understanding many dynamic phenomena in other physical system.

The article is organized as following: In the section II, we
proposed the thermodynamic equation for computing the melting
temperature of dimer film on a constant area. The width of the
rectangle varies from N to N+1. In the section III, we take an
infinitely long dimer film with finite width, and study how the
entropy growth rate behaves when the width increases at small step.
In section IV, we study the entropy difference when two small
rectangles fuse into one big rectangle and when two small torus fuse
into a big torus. In section V, the correlation function of two
loops on torus is computed to check its odd-even phenomena. The last
section is a summary.

\section{The melting temperature of dimer film on finite rectangular lattice with constant area}

The melting temperature of polymer is the temperature at which a
crystalline of polymers transforms into a solid amorphous phase. As
a mathematical modeling of polymer melting transformation, we first
define the frozen phase and the melted phase of dimer film. The
frozen phase is highly ordered phase, all dimers are oriented in the
Y-direction. None of the dimers can rotate to X-direction. The total
number of dimer configuration in the frozen phase is just one. The
melted phase is a disorder phase, the dimer at each lattice site can
transform from a X-oriented dimer into a Y-oriented dimer or vice
versa. All possible different configurations of dimer covering
should be taken into account. The internal energy difference between
the frozen phase and the melted phase is counted by the number of
weak external bond that we assume it binds neighboring dimers. In
the frozen phase, these weak external bond fixed the orientation of
dimers. In the melted phase, these weak external bond is broken by
thermal energy. As this dimer film is the simplest modeling of the
melting process, we assume it is a first order phase transition
since there is still no experimental results of melting two
dimensional dimer film so far. The first order phase transition
requires that the well-known Gibbs free energy is zero at the
transition point, $\Delta G=\Delta U-T\Delta S=0$.

\begin{figure}[htbp]
\centering
\par
\begin{center}
\includegraphics[width=0.47\textwidth]{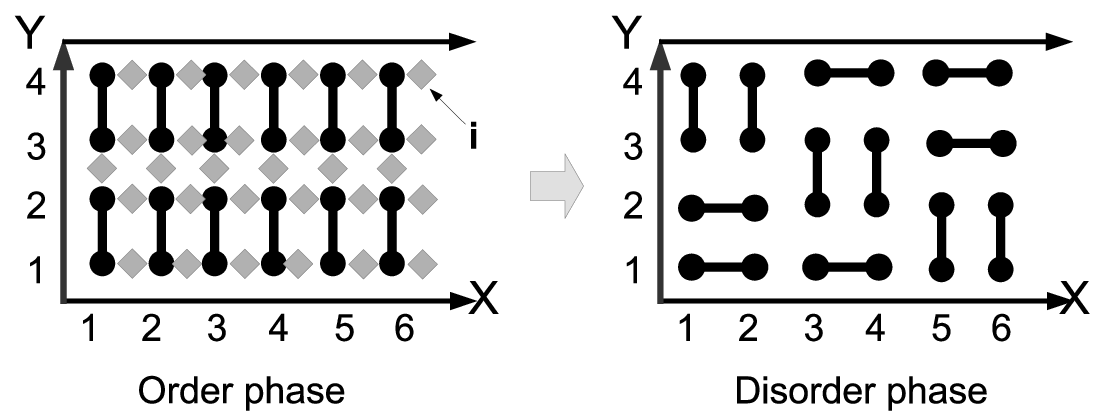}
\end{center}
\caption{\label{film1} The dimer film is approximate infinitely
long. The width varies from $2$ to $16$. The index $i$ indicates the
position of weak external bonds. The grey cross star labels weak
external bonding on the dual lattice of dimer lattice. }
\vspace{-0.2cm}
\end{figure}

The energy quanta for breaking a weak external bond is
$\epsilon_{i}$. The index $i$ labels the position of the weak
external bond on the dual lattice of dimer lattice which is denoted
by the crossing stars in Fig. \ref{film1}. The total thermal energy
for breaking the dual lattice is $\Delta {U}=\sum_{i}\epsilon_{i}$.
Denoting the partition function of the frozen phase as $Z_{1}$ and
the partition function of the melted phase as $Z_{2}$, the melting
temperature reads
\begin{equation}\label{tcc}
T_{c}=\frac{\Delta U}{\Delta S}=\frac{\sum_{i}\epsilon_{i}}{k_{B}
\log[Z_{2}]-k_{B} \log[Z_{1}]} .
\end{equation}
$k_{B}=1.3806488 \times 10^{-23} m^2 Kg s^{-2} K^{-1}$, is the
Boltzmann constant. This equation holds for different lattices in
different dimensions. Here we only consider a two dimensional finite
rectangle of square lattice with m rows and n columns. The number of
lattice sites is $m\times n$. The total number of internal bond and
external bond is $(2mn-n-m)$. The number of internal bonds within
dimer is $mn/2$. Thus the total number of external bonds is
$\frac{3}{2}mn-n-m$. The thermal energy breaks the external bonds to
melt the dime film. If every external bond possesses the same energy
quanta $\epsilon_{0}$, the thermal energy absorbed by the dimer film
at the  temperature is $\Delta U=(\frac{3}{2}mn-n-m)\epsilon_{0}$.

In the frozen phase, the entropy is zero since the number of dimer
configuration is only one. In the melting phase, all different dimer
coverings on the rectangle lattice is possible, thus we apply
Kasteleyn's method \cite{Kasteleyn} to derive an exact counting. The
determinant of Kasteleyn's adjacent matrix $K$ on square lattice
reads \cite{Kasteleyn}
\begin{equation}\label{k}
K=\prod_{k=1}^{n}\prod_{j=1}^{m} (2\cos[\frac{\pi j}{m+1}]+2 i
\cos[\frac{\pi k}{n+1}]).
\end{equation}
Here m and n are both even number. The number of all possible
configurations of dimer covering is $Z=\sqrt{\det K}$
\cite{Kasteleyn}. For other cases, m is odd or even and n is odd or
even, the determinant of Kasteleyn's adjacent matrix $K$ has similar
formulation as Eq. (\ref{k}) \cite{Kasteleyn}. Kasteleyn's method
was mapped into an equivalent free fermion model in quantum field
theory \cite{Samuel} which provides a physical picture of dimer
covering of fermion pairs. Two dimers have a hard-core repulsive
interaction between them since they can not overlap each other.

\begin{figure}[htbp]
\centering
\par
\begin{center}
\includegraphics[width=0.40\textwidth]{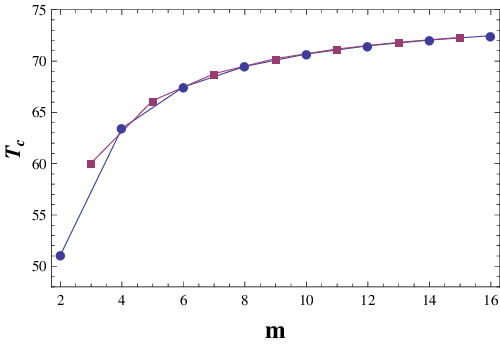}
\end{center}
\caption{\label{tc} The melting temperature of dimer film at
different width from $2$ to 16.  We take the unit energy of an
external bond as $\epsilon_{0}=10^{-4}$ for simplicity.}
\vspace{-0.2cm}
\end{figure}

The odd-even effect is finite size effect, so we choose a finite
square lattice with a large number of lattice sites of
$mn=\Pi_{j=1}^{7}j \times11\times13=720720$. The energy quanta of
weak external bonding is taken as $\epsilon_{0}=10^{-4}$ for the
convenience of computation. In the computation, the total number of
lattice sites is invariant but the ratio of height to width varies.
The width of the long rectangular belt $m$ runs from 2 to 16. The
length of this long rectangle at n=2 is $(720720/2)$, it is almost
infinitely long. The finite size effect is obvious for the width but
is diminished in the lengthy direction. The output of Eq.
(\ref{tcc}) gives the melting temperature. The melting temperature
increases as the width grows from N to N+l(Fig. \ref{tc}). When N
grows larger than $16$, the melting temperature approaches to a
constant value. The odd-even effect is suppressed in thermal dynamic
limit. The temperature curve connecting the odd number of width
floats above that threads the even number of width. There is a
similar but different phenomena in an experimental measurement of
the melting temperature of n-alkanes $C_{n}H_{2n+2}$ powder. The
 temperature curve connecting the odd number of length of
$C_{n}H_{2n+2}$ is below that for the even number of length of
$C_{n}H_{2n+2}$ \cite{Boese}\cite{Hagele}. Even though here we study
the width dependence of dimer film which is different from the three
dimensional experimental system, this theory maybe is still helpful
for an exact understanding on the odd-even effect of polymer
powders.

\begin{figure}[htbp]
\centering
\par
\begin{center}
\includegraphics[width=0.40\textwidth]{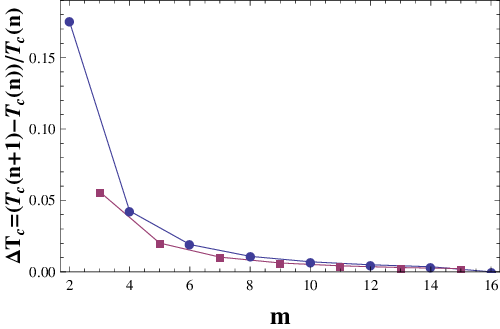}
\end{center}
\caption{\label{dtc} The increase of melting temperature at each
step where the width of the dimer film grows bigger by one unit.}
\vspace{-0.2cm}
\end{figure}

The  temperature increase from a width of N to N+1 also depends on
the odd-evenness of the width number. The temperature increase from
a width of N(N is even number)to N+1 is always bigger than that from
a width of N+1 to N+2. For N=2, the temperature increases by
$0.175617$ percent from a width of 2 to 3. While it only increased
$0.0561494$ percent from a width of $3$ to 4. The temperature
increase curve of $\Delta T_{c}$ that threads the even number of
width is above that passing through the odd number of width (Fig.
\ref{dtc}).

\begin{figure}[htbp]
\centering
\par
\begin{center}
\includegraphics[width=0.40\textwidth]{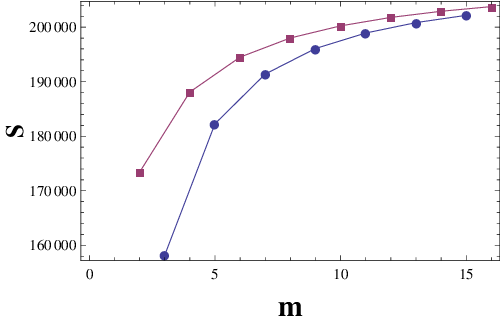}
\end{center}
\caption{\label{polymer} The entropy of dimer coverings on the
infinitely long rectangle. The total lattice sites of the rectangle
is mn=$\Pi_{i=1}^{7}i$ $\times11\times13$=$720720$. The width of
rectangle is represented by $\textit{m}$. } \vspace{-0.2cm}
\end{figure}

The odd-even effect of  melting temperature originates from the
finite size effect of entropy. The entropy curve connecting the even
number of width is above that connecting the odd number of
width(Fig. \ref{polymer}). The entropy increase from m(m is even) to
m+1(m is even) is negative. The entropy increase from from m to m+1
becomes positive for an odd number of m (Fig. \ref{polymer}). If the
long rectangle behaves like a water droplet which can vary its own
shape following the second law of thermodynamics. There would be an
entropy force when the rectangle reshapes itself. We define the
entropy force induced  as $F_{A} =T({\Delta S(m,n)}/{\Delta A}),$
where the area of the film is $A=mn$, $T$ is temperature. The area
is different for different width, so is the entropy covering the
area. By this entropy force, we could predict the interaction of the
film with its environment. Finite size effect determines the
interaction between dimer film and its surrounding.

\section{The  melting temperature of dimer film}

\begin{figure}[htbp]
\centering
\par
\begin{center}
\includegraphics[width=0.40\textwidth]{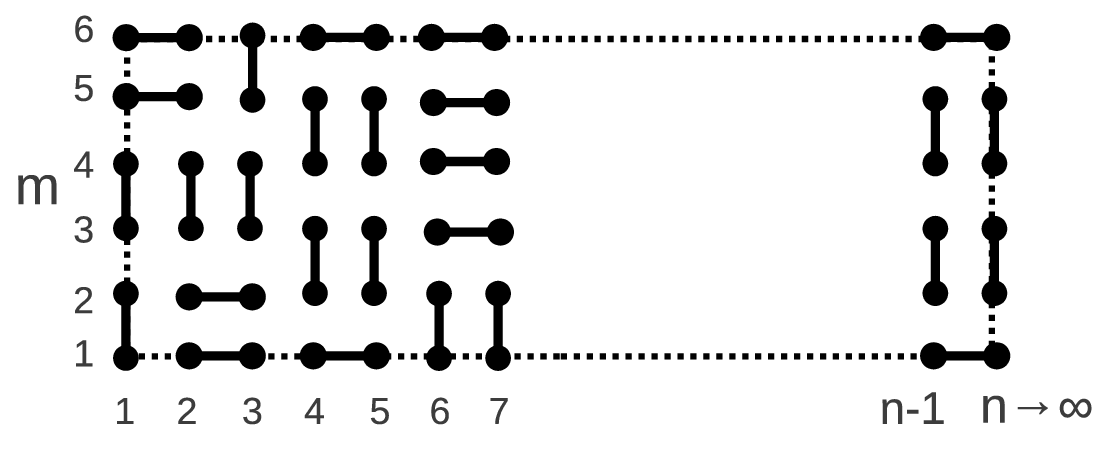}
\end{center}
\caption{\label{film2} The width of the dimer film is finite and is
increasing step by step. The length of the dimer film would be
infinitely long.} \vspace{-0.2cm}
\end{figure}

The melting temperature of dimer film with finite width approaches
to a constant value in thermal dynamic limit. To explicitly express
that constant value we study the entropy growth rate of an
infinitely long rectangle when its length increases step by step
(Fig. \ref{film2}). The number of rows is fixed to a small value.
The entropy of finite system is defined as the same formulation as
thermodynamic entropy, $S=k_{B}\log[Z]$. Here Boltzmann constant
$k_{B}$ is set to $1$ for convenience. For fixed number of rows, the
entropy growth for a growing number of columns by one unit reads
\begin{equation}
\Delta{S}_{m,n}=k_{B}\log{[Z(m,n)]}-k_{B}\log{[Z(m,n-1)]}.
\end{equation}
The  melting temperature is determined by the ratio of internal
energy difference to entropy difference, $T_{c}={\Delta U}/{\Delta
S}$. So entropy difference can be deduced by the melting temperature
equation, ${\Delta S}={\Delta U}/ T_{c}$. The internal energy
difference between n column and (n-1) column is $\Delta
U=(\frac{3}{2}mn-n-m)\epsilon_{0}=N_{m,n}\epsilon_{0}$. The
combination of  $\Delta U$ and ${\Delta S}$ gives the entropy growth
from $n$ column to $(n-1)$ column for a fixed number of $m$ rows,
\begin{equation}\label{dsmn}
\Delta{S}_{m,n}=N_{m,n}(\frac{\epsilon_{0}}{T_{m,n}}-\frac{\epsilon_{0}}{T_{m,n-1}})+(\frac{3}{2}m-1)\frac{\epsilon_{0}}{T_{m,n-1}}.
\end{equation}
According to the computation of melting temperature of long belt in
last section, the temperature difference of
$\Delta{T}_{c}=T_{c}(n)-T_{c}(n-1)$ for large number of $n$
approaches to zero. The first term at the right hand side of Eq.
(\ref{dsmn}) approaches to zero when the number of columns goes to
infinity,
$\lim_{n\rightarrow\infty}N_{m,n}({\epsilon_{0}}/{T_{m,n}}-{\epsilon_{0}}/{T_{m,n-1}})=0$.
As a result, the entropy increase now only depends on the number of
rows,
\begin{equation}\label{dsm}
\Delta{S}_{m,n\rightarrow\infty}=(\frac{3}{2}m-1)\frac{\epsilon_{0}}{T_{m,n\rightarrow\infty}}.
\end{equation}
The right hand side of Eq. (\ref{dsm}) is actually the derivative of
entropy with respect to the number of columns in thermal dynamic
limit. If we take a further derivative of
$\Delta{S}_{m,n\rightarrow\infty}$ with respect to the number of
rows $m$, it leads to constant $\kappa$,
\begin{eqnarray}
\kappa=\frac{\partial}{\partial {m}}\left
[\lim_{n\rightarrow\infty}\frac{\Delta{S}}{\Delta{n}}\right]=\frac{3}{2}\frac{\epsilon_{0}}{T_{m,n\rightarrow\infty}}.
\end{eqnarray}
As $(\Delta{m}\Delta{n})$ indicates the area of an unit square on
square lattice. $k$ is in fact the entropy per square. One square
occupies four quarters of one lattice site. $k$ is also the entropy
per site. The constant melting temperature in thermal dynamic limit
now has a brief formulation,
\begin{eqnarray}\label{tcinfty}
T_{n\rightarrow\infty}=\frac{3}{2}\frac{\epsilon_{0}}{\kappa}.
\end{eqnarray}
This equation reveals the relationship between the  melting
temperature, entropy per site and external bonding energy. A
stronger external bonding hold dimers together with more strength,
thus it takes more thermal energy to melt. Melting process increases
the entropy of dimer film to make it more chaos, so it also costs
higher thermal energy to melt a highly ordered film with less
entropy.

\begin{figure}[htbp]
\centering
\par
\begin{center}
$
\begin{array}{c@{\hspace{0.03in}}c}
\includegraphics[width=0.40\textwidth]{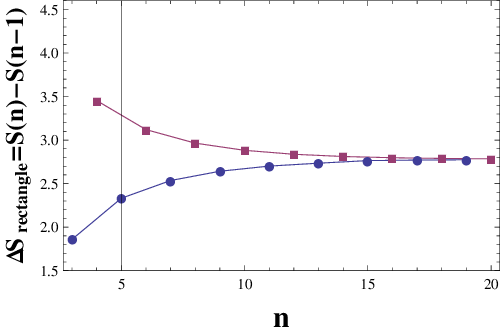}& \\
\mbox{(a)} &\\
\includegraphics[width=0.40\textwidth]{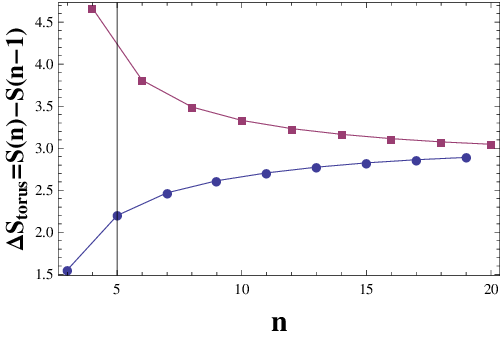}&\\
\mbox{(b)} & \\
\end{array}
$
\end{center}
\caption{\label{openeo} (a) The entropy growth of a rectangle from odd number of columns to even number of columns, or vice verse. The Number of rows is
fixed, m=10.
(b)The entropy growth of a torus. The number of rows is kept to m=10. The number of columns varies from odd to even, or vice verse. } \vspace{-0.2cm}
\end{figure}

\begin{figure}[htbp]
\centering
\par
\begin{center}
\includegraphics[width=0.40\textwidth]{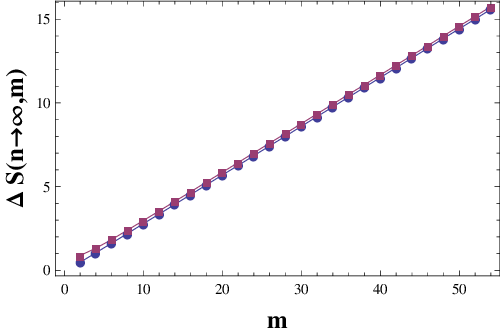}
\end{center}
\caption{\label{openinf} The limit value of entropy growth for $n\rightarrow\infty$ at different number of rows $m$. The square dot represents torus.
The disc dot represents rectangle. The limit value of torus is always higher than that of rectangle. } \vspace{-0.2cm}
\end{figure}

The external bonding energy can only be determined by experimental
measurement. The entropy per site in the liquid phase of dimer film
can be obtained by Kasteleyn's method. Here we derive the entropy
per site by a different approach of Eq. (\ref{dsm}). The linear
dependence of $\Delta{S}_{m,n\rightarrow\infty}$ on the number of
rows $m$ is verified by direct computation. We first computed the
entropy growth rate for finite length and then extended the
computation to infinite length. The computation is performed both
for a rectangle and a torus. A rectangle has open boundary
condition. A torus has periodic boundary condition in both the row's
direction and the column's direction. Fig. \ref{openeo} shows the
entropy growth on a rectangle and a torus when the number of column
increases from 2 to 20. The number of rows is fixed to $m=10$. The
entropy growth for the length growing from $n-1$ to $n$(n is even)
is always bigger than the case that n is odd (Fig. \ref{openeo}).
The entropy growth curve for an odd number of length $n$ increases
as the length becomes longer. While the entropy growth curve
connecting those even number of length decays. These two curves
converges to the same constant value when the length of rectangle
becomes bigger than 20. This phenomena also holds for torus, the
convergence point of the two curves for torus is bigger than that of
rectangle. It takes more steps for $\Delta{S}$ to converge on a
torus than that on a rectangle. The limit value
$\Delta{S(n\rightarrow\infty,m=10)}$ on rectangle is
$\Delta{S_{rec}(n\rightarrow\infty,m=10)}=2.778441$, here we have
made a cut off at $10^{-6}$. The limit value on a torus for large
$n$ is a little bit higher,
$\Delta{S_{tor}(n\rightarrow\infty,m=10)}=2.969359$. The difference
between the limit values of rectangle and torus vanishes for
$m\rightarrow\infty$. Fig. \ref{openinf} showed the limit values of
$\Delta{S_{tor}(n\rightarrow\infty)}$ and
$\Delta{S_{rec}(n\rightarrow\infty)}$ when $\textit{m}$ grows from 2
to 54. Both the two limit values form a straight line.

The limit values fit well with an empirical linear equation for
large $\textit{m}$($m>100)$,
\begin{eqnarray}
\Delta{S_{rec}(n\rightarrow\infty,m)}&=&0.29\;m+C_{rec}(m),\nonumber\\
\Delta{S_{tor}(n\rightarrow\infty,m)}&=&0.29\;m+C_{tor}(m), \nonumber\\
 m&=&2k,\;\;k=0,1,2,3,\cdots.
\end{eqnarray}
The accuracy of the numerical values in equations are kept to the
order of $10^{-2}$ for brevity without losing the key features. This
empirical linear equation implies a constant slope,
\begin{eqnarray}
\kappa/k_{B}=\frac{\partial}{\partial {m}}\left
[\lim_{n\rightarrow\infty}\frac{\Delta{S}}{\Delta{n}}\right]=0.29.
\end{eqnarray}
This is consistent with the computed  temperature in Fig. (\ref{tc})
where $T_{c}$ approaches to a constant for large $m$. The
probability of a dimer on square lattice obtained by Kasteleyn is
1.791622 \cite{Kasteleyn}. The entropy per site reads
0.29=log(1.791622)/2. This number is exactly the same as the slope
of $\Delta{S_{tor}}(n\rightarrow\infty)$ for large m. According to
the melting temperature equation (\ref{tcinfty}), the physical
entropy per site for a dimer film on square lattice is
$\kappa=0.29\times k_{B}=0.400388152 \times 10^{-23} m^2 Kg s^{-2}
K^{-1}$. The physical process of melting a real liquid crystal film
experiences a multi-step transitions from an ordered phase into a
liquid phase \cite{Chou}. For instance, the liquid crystal film
transform a crystal-B phase into a hexatic-B phase at 60 $^\circ${C}
\cite{Chou}. As the temperature raises to 63 $^\circ${C}, the liquid
crystal film transform stepped into Sm-A phase \cite{Chou}. It melts
into an ordinary two dimensional liquid at about 66.3 $^\circ${C}
\cite{Chou}. The melting temperature and external bonding energy
quanta in Eq. (\ref{tcinfty}) is not a quantity that can be
determined by theoretical computations only, it can only be
determined by experimental measurement. As this model does not focus
on specific molecules,  Eq. (\ref{tcinfty}) also holds for liquid
crystal film. Thus we simply assume the critical melting temperature
is roughly 67 $^\circ${C}, i.e., $T_{c}=$ 340K. Then we could
estimate the external bonding energy,
\begin{eqnarray}\label{bondenergy}
\epsilon_{0}=\frac{2}{3}{\kappa} T_{c}=0.907546778\times 10^{-21}
m^2 Kg s^{-2}.
\end{eqnarray}
Thus the weak external bonding energy is about $0.907546778\times
10^{-21}J$. As all know, $1 eV = 1.602176565(35)\times 10^{-19} J$,
thus the weak external bonding energy is about $\epsilon_{0}=
0.566446169\times 10^{-2}$ev. While the usual Carbon-Carbon bond
energy is around 3.60$\sim$3.69 ev. Since $\epsilon_{0}$ is much
smaller than the chemical bond, we call it weak external bond.

The difference between $\Delta{S_{tor}}(n\rightarrow\infty)$ and
$\Delta{S_{rec}}(n\rightarrow\infty)$ is a function of $\textit{m}$,
$\Delta{C(m)}=C_{tor}(m)-C_{rec}(m)$. The maximal value of
$\Delta{C(m)}$ is $\Delta{C(2)}$=0.400162. $\Delta{C(m)}$ decays
when $m$ grows. The most rapid decay occurs from $m=2$ to $m=50$.
When $\textit{m}$ continue to grow from $m=50$ to $m=500$,
$\Delta{C(m)}$ only drops from $0.157040$ to $0.149912$.

\begin{figure}[htbp]
\centering
\par
\begin{center}
\includegraphics[width=0.40\textwidth]{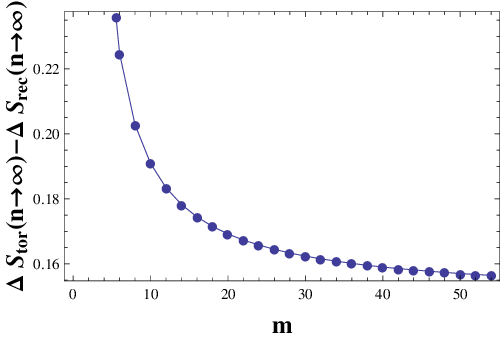}
\end{center}
\caption{\label{gapinf} The value of $\Delta
C(m)=\Delta{S_{tor}(n\rightarrow\infty)}-\Delta{S_{rec}(n\rightarrow\infty)}$
as a function of the number of rows $\textit{m}$. $\Delta C(m)$
approaches zero as $m$ goes to infinity.} \vspace{-0.2cm}
\end{figure}

The topological difference between rectangle and torus can be characterized by their different speed of entropy growth,
\begin{equation}\label{toponumber}
\Delta C(m)=\Delta{S_{tor}(n\rightarrow\infty)}-\Delta{S_{rec}(n\rightarrow\infty)}.
\end{equation}
The rapid decay of $\Delta C(m)$ is illustrated in Fig.
\ref{gapinf}. A logarithmic plot of $\Delta C(m)$ with respect to m
still behaves as rapid decay instead of a straight line. The
decaying rate of $\Delta C(m)$ is much faster than exponential
decay. The topological difference between torus and rectangle
vanishes when the width goes to infinity. The melting temperature in
thermal dynamic limit does not depend on boundary condition.

\section{The entropy principal of fusing two small dimer films into one big dimer film}

The second law of thermodynamics requires that fusing two different
types of molecule into one solution would increase the entropy of
the system. Two molecules would dissolve into each other much easier
if the entropy increased much more. The odd even effect can be
understand by the second law of thermodynamics. First, we calculate
the entropy before the two molecules meet, $S_{a}$. Then, calculate
the entropy after they dissolve in each other, $S_{b}$. The entropy
difference in this process, $\Delta{S}=S_{b}-S_{a}$, determines
wether they dissolve into each other or not. The practical liquid
crystal molecule is too complicate to accomplish an exact
calculation of entropy. We still focus on the ideal dimer film to
check if there exist odd even effect when two small films fuse into
a big film(Fig. 2).

\begin{figure}[htbp]
\centering
\par
\begin{center}
\includegraphics[width=0.35\textwidth]{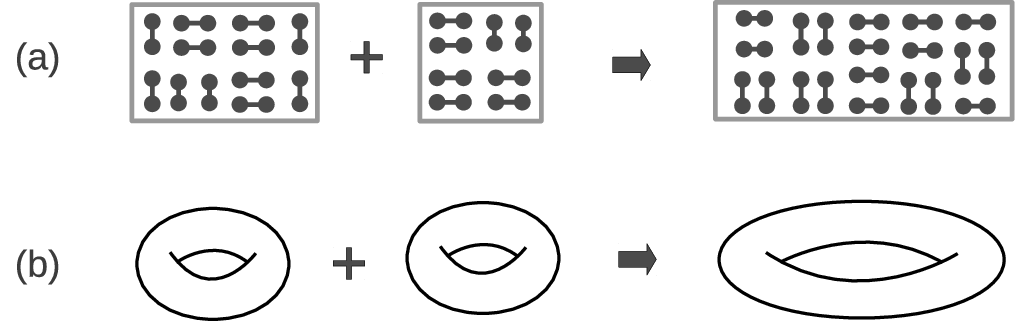}
\end{center}
\caption{\label{fusion} (a) Fusing two small rectangles with finite
size into one big rectangle. (b) Fusing two small toruses into one
big torus. We compute how much the entropy increased after the
fusion.} \vspace{-0.2cm}
\end{figure}

The entropy is a monotonic logarithmic function of the number of
configurations. We use the number of configurations to quantify the
entropy of dimer film. The rectangle of square lattice has m rows
and n columns. The number of rows $m$ is fixed in computation. Only
the number of columns $n$ is decomposed as the sum of two smaller
numbers: $p$ and $n-p$.  The increase rate of possible dimer
configurations for fusing two separated film into one is
\begin{eqnarray}
{\Delta}Z=\frac{Z_{m,n}-Z_{m,p}Z_{m,n-p}}{Z_{m,n}}.
\end{eqnarray}
Because ${\Delta}Z$ is not always positive, if we take the logarithm
function of ${\Delta}Z$ to compute the entropy, the information of
negative ${\Delta}Z$ would be lost. The number of dimer
configurations increases(decrease) as entropy increase(decrease).
${\Delta}Z$ can quantify the variation of entropy.

\begin{figure}[htbp]
\centering
\par
\begin{center}
$
\begin{array}{c@{\hspace{0.03in}}c}
\includegraphics[width=0.35\textwidth]{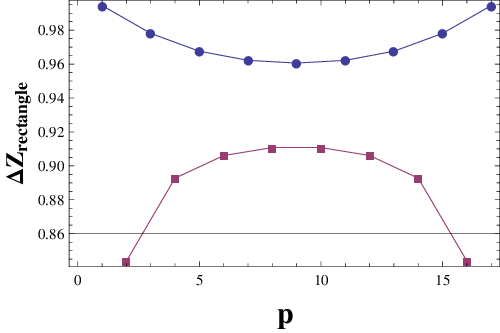}& \\
\mbox{(a)}  &\\
\includegraphics[width=0.35\textwidth]{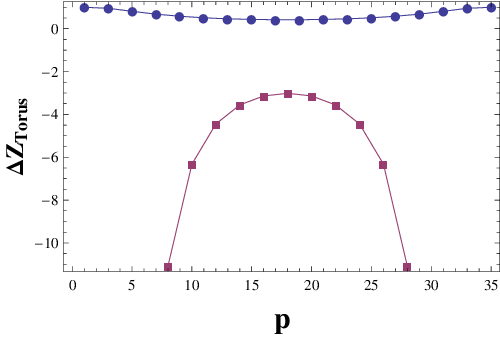}&\\
\mbox{(b)}  & \\
\end{array}
$
\end{center}
\caption{\label{recut} (a) The increased number of states after
dividing a finite rectangle into two smaller parts. The big
rectangle has $\textit{m}=18$ rows and $\textit{n}=18$ columns.
$\textit{m}=18$ is invariant. The length of $\textit{n}$ is divided
into two parts, such as (1,17), (2,16), (3,15), $\cdots$, and so
forth. (b)The difference of total number states before and after
dividing a finite torus with $\textit{m}=\textit{n}=36$. }
\vspace{-0.2cm}
\end{figure}

We took a sequence of pair of two small dimer film with the number
of columns, $\{(1,n-1), (2,n-2), (3,n-3),\cdots\}$. The two small
films fuse into a big film with $n$ columns. The entropy increase
for fusing two dimer films of an odd length is larger than that for
fusing two dimer films with an even length(Fig. \ref{recut} (a)).
The entropy gain forms two bands for both rectangle and torus. The
upper band is the entropy increase at odd numbers. The lower band is
the entropy increase at even number of length. There is a
significant difference between rectangle and torus. For rectangle,
the entropy increase is positive for both fusing two odd small
rectangles and two even rectangles. For torus, the entropy increase
for fusing two odd small torus is positive. However fusing two even
torus does not increase entropy, instead the entropy is reduced
after the fusion (Fig. \ref{recut} (b)). According to the second law
of thermodynamics, the fusion of two odd rectangles or torus is most
favored by entropy. Two even rectangle can fuse into one, but two
even torus would repel each other. Topological difference is
significant in finite size scale. We doubled the length of the
mother torus by keeping $m$ invariant. The gap of the entropy gain
between even number and odd number becomes smaller(Fig. 12).

For the two bands of entropy increase, the maximal variation occurs
when the two small dimer films is significantly different in length.
The minimal entropy increase is situated at the middle point where
the two small dimer films have almost identical length. The two
spatial dimension parameters play competing roles in controlling the
gap between the two bands. If we keep $n$ invariant and doubled $m$,
the entropy increase for fusing two odd torus approximate to zero.
But entropy reduction for fusing two even torus becomes stronger.
The band gap between even and odd grows wider. Therefore increasing
$n$ makes smaller band gap, while increasing $m$ enlarge the band
gap.

\begin{figure}[htbp]
\centering
\par
\begin{center}
$
\begin{array}{c@{\hspace{0.03in}}c}
\includegraphics[width=0.35\textwidth]{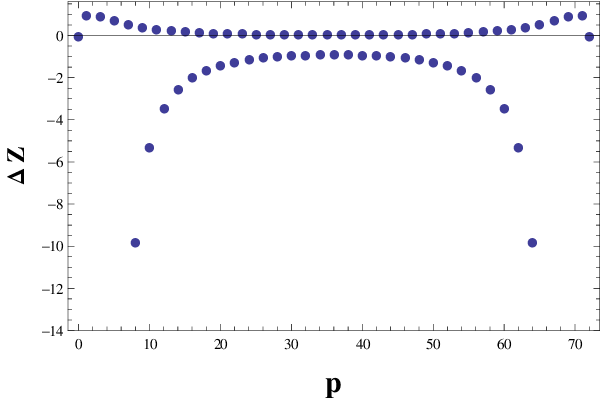}& \\
\mbox{(a)}    &\\
\includegraphics[width=0.35\textwidth]{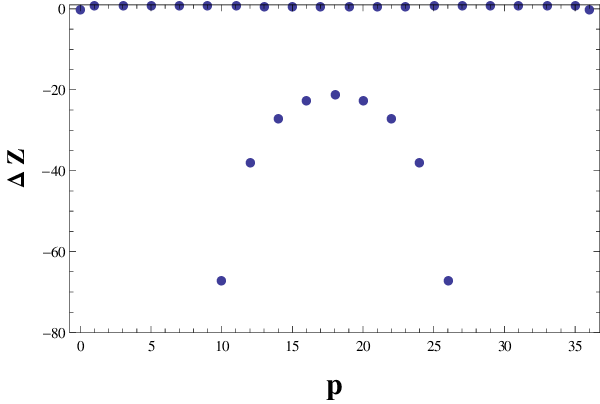}&\\
\mbox{(b)  } & \\
\end{array}
$
\end{center}
\caption{\label{m72n36} (a) The variation of number of states
for dividing a finite torus with ($\textit{m}=72$, $\textit{n}=36$) and uniting them.
Comparing with the finite torus in Fig. \ref{recut} (b),
the number of rows is doubled.
(b) The difference of number of states before and after dividing the finite torus with
($\textit{m}=36$, $\textit{n}=72$). The number of columns is doubled comparing with Fig. \ref{recut} (b). } \vspace{-0.2cm}
\end{figure}

According to the second law of thermodynamic, we may come to the
conclusion only two odd toruses of dimer film can fuse into one big
torus. If a big torus splits into two small torus, it is most likely
two even torus. Fusing two odd rectangles is easier than fusing two
even rectangles. This theoretical study maybe is helpful for
understanding why two liquid crystal molecules are miscible only
when the linker includes odd number of units.

\section{The correlation of two intersecting loops on torus}

\begin{figure}[htbp]
\centering
\par
\begin{center}
\includegraphics[width=0.42\textwidth]{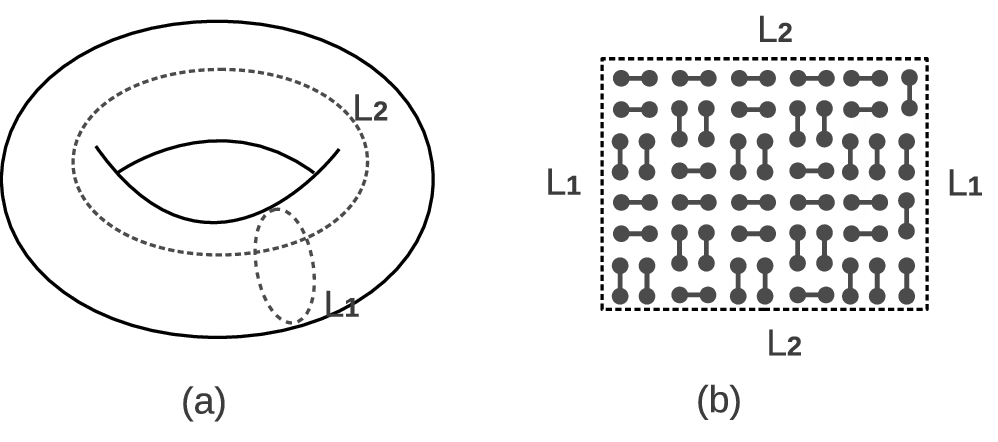}
\end{center}
\caption{\label{torus} (a) A torus with two topologically
inequivalent circles, $L_{1}$ and $L_{2}$. (2) A torus covered by
dimers can be constructed from a rectangle lattice by taking
periodic boundary condition in two spatial dimensions. }
\vspace{-0.2cm}
\end{figure}

Torus is a manifold generated by sweeping one circle around another
circle(Fig. \ref{torus} (a)). A torus is equivalent to a rectangle
with periodic boundary condition in both of the two spatial
dimensions. The two circles, $L_{1}$ and $L_{2}$, are the boundaries
of rectangle(Fig. \ref{torus} (b)). The entropy of dimer film is
reduced when a torus film transforms into a rectangle film. The
entropy reduction during this topological transformation quantifies
the geometric correlation between $L_{1}$ and $L_{2}$.

In the free fermion theory of dimer model, the correlation of two
monomers is defined by the ratio of the total number of
configurations with two monomers to the case without two monomers
\cite{Moessner}\cite{Fendley}\cite{tzeng}. The two loops are long
string of many monomers. Usually the number of dimer configurations
of rectangle is smaller than that of torus. For a finite lattice,
the difference of the number of configurations between a torus and a
rectangle is small, the subtle information of loop correlation is
almost invisible by the definition of the references above. Here we
act a logarithm operation on the traditional correlation function to
extract out the subtle odd-even effect. The correlation of the two
intersecting loops is defined as
\begin{eqnarray}
C(L_{1},L_{2})&=&\log[\langle{\psi_{1}\psi_{2}\cdots\psi_{n}\psi_{o}\psi_{1}\psi_{2}\cdots\psi_{m}}\rangle],\nonumber\\
&=&\log[\langle{L_{1}L_{2}}\rangle]
\end{eqnarray}
Loop $L_{1}$ is a string of $n+1$ monomers,
$\psi_{1}\psi_{2}\cdots\psi_{n}\psi_{o}$. Loop $L_{2}$ covers $m$
monomers, $\psi_{1}\psi_{2}\cdots\psi_{m}$, and the monomer at the
intersecting point, $\psi_{o}$. This loop correlation function
quantifies the correlation among $(n+m+1)$ fermions. Here
$C(L_{1},L_{2})$ is actually the entropy difference between
rectangle and torus,
\begin{eqnarray}
C(L_{1},L_{2})=\log[\frac{Z_{rec}}{Z_{tor}}]=S_{rec}-S_{tor}.
\end{eqnarray}
We keep the total number of fermions expanding the two loops as
constant, $L_{1}+L_{2}=n+m+1=const$. Generally speaking, the
geometric correlation between the two loops is negative. If one
circle becomes larger, the other circle must be smaller.

The correlation of two loops shows odd-even dependence on the length
of the loop(Fig. 14). First, we keep $m+n=48$, and both ($m$,$n$)
are even. The correlation is weak at small $n$ and large $n$, the
maximal correlation is around the middle, $n=24$. Then, we still
keep $m+n=48$, but modify $n$ and $m$ in a sequence of odd numbers.
The strong correlation now appears at small $n$ and large $n$.
Around $n=23$ is the minimal correlation which is still larger than
the maximal value for even case. We come to the conclusion the
correlation between two loops with odd number of length is stronger
than the correlation of loops with even number of length. The
correlation has a finite gap between the odd and the even case. When
the length of the two loops grows to infinity, this gap becomes
zero. The odd even effect of loop correlation holds for other hybrid
cases: (1) m is even and n is odd. The correlation curve inherits
one half of the even case and one half of the odd case. For
$m+n=49$, the maximal correlation appear at small $n$, and the
minimal correlation appears at large n. (2) m is odd and n is
even(Fig. 14). For $m+n=49$, the correlation curve is the mirror
image of case (1) reflected by the vertical axis $n=23$.

\begin{figure}[htbp]
\centering
\par
\begin{center}
$
\begin{array}{c@{\hspace{0.03in}}c}
\includegraphics[width=0.35\textwidth]{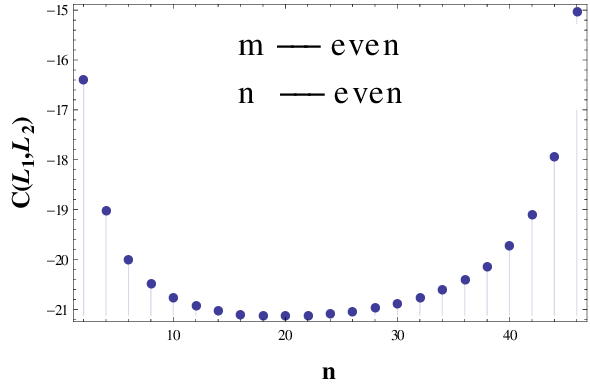}& \\
\mbox{(a)}&\\
\includegraphics[width=0.35\textwidth]{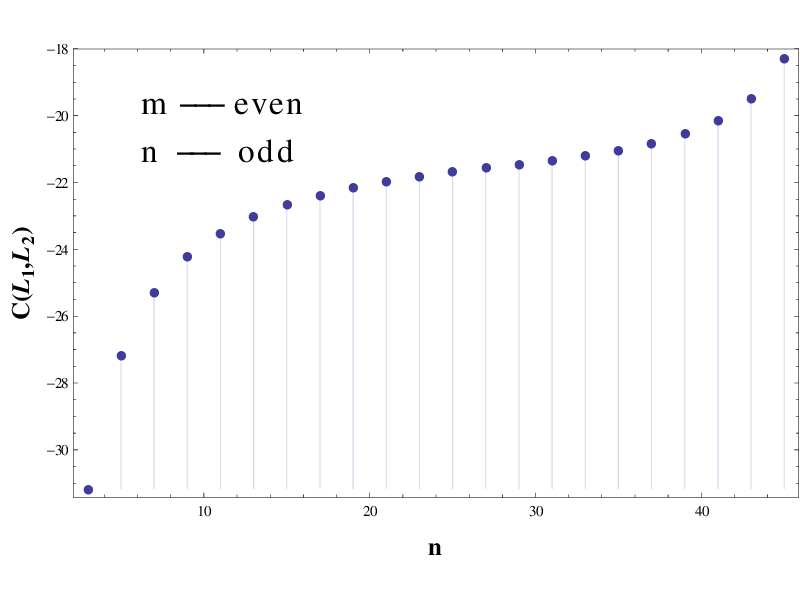}&\\
\mbox{(b)} & \\
\includegraphics[width=0.35\textwidth]{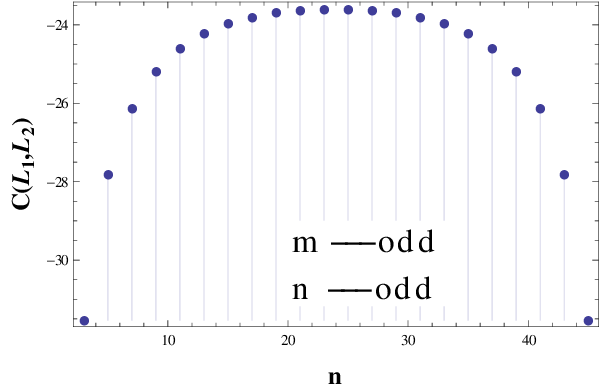}&\\
\mbox{(c)} & \\
\end{array}
$
\end{center}
\caption{\label{meno} (a)The sum of the two loops is constant,
$L_{1}+L_{2}=49$, i.e., m+n=48. Both $\textit{m}$ and $\textit{n}$
are even number varying from 2 to 48. (b) The total length of the
two loop is 50, m+n=49. The length of $L_{2}$ varies as following,
(2,4,6,$\cdots$,48). The corresponding length of $L_{2}$ is (47, 45,
43,$\cdots$,3,1). (c) The sum of the two loop is 49, m+n=48. Both
$\textit{m}$ and $\textit{n}$ varies from odd number to odd number.}
\vspace{-0.2cm}
\end{figure}

The loop correlation function $C(L_{1},L_{2})$ define the geometric
correlation of the two finite loops. The two free fermion loops also
obey a topological constraint as long as the fermion loop does not
break into an open chain.

\section{The melting temperature of long polymer's film}

\begin{figure}[htbp]
\centering
\par
\begin{center}
\includegraphics[width=0.42\textwidth]{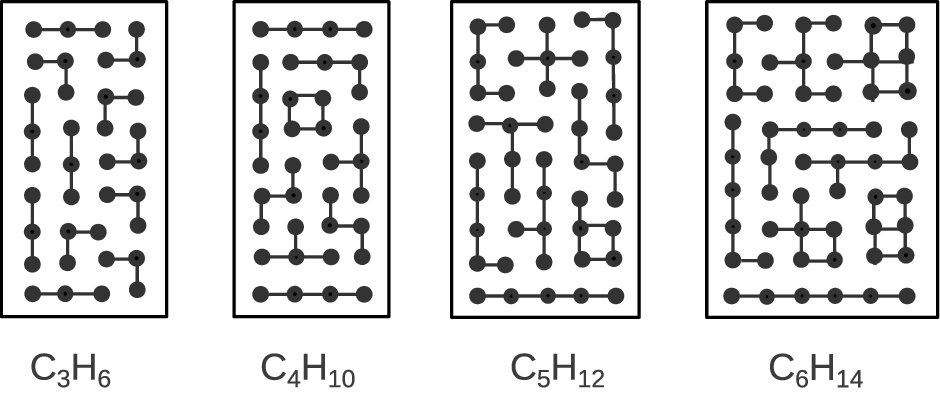}
\end{center}
\caption{\label{trimer} Covering of a finite square lattice by long
polymers. The length of the mathematical polymer showed above is 3,
4, 5, 6. The mathematical polymer represents the n-alkanes
$C_{n}H_{2n+2}$. Soft polymer can bend into different elementary
configurations. } \vspace{-0.2cm}
\end{figure}

The melting temperature for a film of long polymers can be computed
by the same critical temperature Eq. (\ref{tcc}) for dimers. First
we need to count the number of external bonds. Then we count all
possible ways to cover the lattice by trimer(3-connected monomers),
quadramer(4-connected monomers), pentagomer(5-connected monomers),
and so on(Fig. \ref{trimer}). However it is a difficult mathematical
problem to exactly count all possible polymer coverings by long
polymers. So far we only have the exact solution of dimer covering
problem. For the simplest case that the polymer is rigid enough to
keep the shape of a straight line, the entropy of longer polymer
would be smaller than the entropy of shorter polymers for covering
the same area. We can always cut the longer polymers shorter, as a
result, the number of possible configurations would increase.
According to Eq. (\ref{tcc}), smaller entropy induces larger melting
temperature. Although the number of external bonds also decreases
for longer polymer, it falls far behind the decreasing speed of
entropy. Qualitatively we can predict that the melting temperature
would increase when the length of the polymer increases. In mind of
the experimental observation of odd-even effect, we believe the
exact entropy of long polymer covering would demonstrate odd-even
effect. In fact, it is possible to measure the melting temperature
of a film in lab. So experiment maybe is another way for solving
mathematical problems.

In reality, the n-alkanes $C_{n}H_{2n+2}$ is not very rigid for
large $n$. The soft chain would bend into different
configurations(Fig. \ref{trimer}). In that case, the entropy of
longer polymer maybe is larger than shorter polymers. Then we would
meet the anomalous odd-even effect. For a more complicate case, if
we mix polymers of different length, it is almost impossible to get
exact mathematical counting of all possible coverings. But we can
divide the polymers into two classes: even-polymer and odd-polymers.
The odd-even effect would still play a role.

\section{Summary}

Odd-even effect is a common phenomena in finite size system. If
certain degree of freedom of a system is highly confined in finite
scale, the physical observable would strongly depends on the
confined dimensions. We study how the  melting temperature and
entropy depend on the finite width of a long dimer film. As a
mathematical modeling of the melting process of a dimer film, we
assumed there is a weak external bonding between neighboring dimers
and defined the transition between a solid dimer phase and a liquid
dimer phase. Kasteleyn's method was used for computing the entropy
of the liquid dimer phase. The melting temperature curve connecting
the odd width is above that connecting the even width. While the
entropy curve threading the even width is above that of the odd
width. The odd even effect also exist in the fusion of two smaller
dimer films into a bigger dimer film. The entropy increase for
fusing two odd rectangular film is bigger than that for fusing two
even rectangular film. For torus film, only fusing two torus films
with odd number of length increase the entropy. Fusing two even
torus films into a big torus film experiences an entropy decrease.
This odd-even effect is consistent with the second law of
thermodynamics. The entropy of a system tends to increase during a
physical process. The correlation function of two loops on torus
also depends on the odd-eveness of their length. The correlation of
two odd loops is stronger than the correlation of two even loops.

The odd even effect is diminished when the finite width grows longer
until thermal dynamic limit. The melting temperature approaches to a
constant value. We derived the equation of  melting temperature in
thermodynamic limit by studying a growing dimer film. The  melting
temperature is proportional to the external bonding energy and the
inverse of entropy per site. This equation has no dependence on
boundary condition. The difference between a torus and a rectangle
decays to zero as the width goes to infinity. The general equation
for computing  melting temperature of dimer film also hold for
longer polymer, such as trimer, tetramer, and so on. However how to
exactly count all the possible configurations of trimer covering or
tetramer covering on square lattice is an unsolved mathematical
problem. We conjecture that the melting temperature computed by an
extended dimer film model would still demonstrate odd-even
dependence on the length of polymers. There might be a significant
difference between two dimensional dimer film model and three
dimensional dimer powder model. Further study will be performed in
the future.

\section{Acknowledgment}

This work is supported by the Fundamental Research Foundation for
the Central Universities and  National Natural Science Foundation of
China(11304062).

\end{document}